\documentclass[12pt]{article}

\usepackage{amsfonts,amsmath,amssymb}

\usepackage{tikz}

\newcommand{\ra}{\rightarrow}
\newcommand{\Hom}{{\rm Hom}}

\newcommand{\ZZ}{{\mathbb Z}}
\newcommand{\RR}{{\mathbb R}}

\newcommand{\FF}{{\mathbb F}}

\newcommand{\hG}{{\hat G}}
\newcommand{\frP}{{\mathfrak P}}

\newcommand{\tc}{{\tilde c}}

\newcommand{\cH}{{\mathcal H}}

\newtheorem{q}{Question}

\title{Anomalies of discrete symmetries in various dimensions and group cohomology}

\author{Anton Kapustin \\ {\it  California Institute of Technology, Pasadena, CA}\\ Ryan Thorngren \\ {\it University of California, Berkeley, CA}}

\begin{document}

\maketitle

\begin{abstract}

We study 't Hooft anomalies for discrete global symmetries in bosonic theories in $2$, $3$ and $4$ dimensions. We show that such anomalies may arise in gauge theories with topological terms in the action, if the total symmetry group is a nontrivial extension of the global symmetry by the gauge symmetry.  Sometimes the 't Hooft anomaly for a $d$-dimensional theory with a global symmetry $G$ can be canceled by anomaly inflow from a $(d+1)$-dimensional topological gauge theory with gauge group $G$. Such $d$-dimensional theories can live on the surfaces of Symmetry Protected Topological Phases. We also give examples of theories with more severe 't Hooft anomalies which cannot be canceled in this way. \end{abstract}

\section{Introduction}

It is well-known that in even space-time dimensions there can be obstructions to gauging a global symmetry \cite{GrossJackiw}. Such obstructions are  known as 't Hooft anomalies \cite{tHooft} and they are well understood in the case when the symmetry group $G$ is a compact connected semi-simple Lie group (see e.g. ch. 22 of \cite{Weinberg} and references therein). 't Hooft anomaly in such theories typically manifests itself as a lack of gauge-invariance of the effective action for fermions in a background $G$ gauge field. Such an anomaly arises only when the space-time dimension $d$ is even, because it requires a chiral action of $G$ on the fermions. The anomaly is always a phase which depends both on the background gauge field and the gauge transformation. It can be described most easily by saying that it is equal to a boundary term in the gauge variation of the Chern-Simons action for $G$ in $d+1$ dimensions. We express this by saying that the 't Hooft anomaly can be canceled by anomaly inflow from one dimension higher. Since Chern-Simons theories in $d+1$ dimensions are classified by elements of $H^{d+2}(BG,\ZZ)$ \cite{DW}, this means that the 't Hooft anomaly for compact connected semi-simple Lie groups takes values in $H^{d+2}(BG,\ZZ)$. Here and below $BG$ denotes the classifying space of principal $G$-bundles. 

The case of a discrete symmetry $G$ has been studied much less. In a theory of free fermions or a small deformation thereof, one can compute the 't Hooft anomaly by reducing to the case of a compact connected Lie group. In such theories anomalies arise only from the chiral action of $G$ on the fermions. Since the action of $G$ on fermions is linear and unitary, $G$ comes with a distinguished embedding into $U(N)$. The anomaly can be computed by evaluating the $U(N)$ anomaly and then restricting it to $G$. The resulting anomaly takes values in $H^{d+2}(BG,\ZZ)\simeq H^{d+1}(BG,U(1))$ and can be canceled by anomaly inflow from $d+1$ dimensions, where the $d+1$-dimensional topological gauge theory is of the Dijkgraaf-Witten type \cite{DW}. By construction, anomalies in theories of weakly-coupled fermions occur only for even $d$, because only in even $d$ one can have a chiral action of $G$ on fermions. 

Recently anomalies for discrete symmetries have attracted renewed attention \cite{Wen, WangWen,KTshort,Kapustin} mostly because of their importance in the theory of Symmetry Protected Topological Phases (SPT phases). Namely, it has been argued that the surface of a nontrivial SPT phase with symmetry $G$ must be described by a theory with an 't Hooft anomaly for $G$. Examples of SPT phases and their surfaces constructed in the literature \cite{VS}-\cite{decorated} show that 't Hooft anomalies are not restricted to even dimensions and can arise in purely bosonic systems.

In this paper (some of whose results were announced in \cite{KTshort}) we investigate systematically 't Hooft anomalies for discrete global symmetries in bosonic theories. Specifically, the theories we consider are abelian gauge theories in dimension $d$ which may have topological terms in the action. The gauge group $D$ can be either discrete or continuous. The key observation \cite{VS} is that anomalies often arise when the total symmetry group is not a product $G\times D$ but an extension $\hG$ of $G$ by $D$. This means that $D$ is a normal subgroup of $\hG$, and $G=\hG/D$. In many of our examples both $D$ and $G$ are abelian, but $\hG$ is often nonabelian. 

't Hooft anomaly in bosonic gauge theories can exist in both even and odd space-time dimensions. Instead of chirality, anomaly is caused by the conflict between the way $G$ acts on electric and magnetic excitations, or by the conflict between the action of $G$ on electric excitations and topological terms in the action. 

In some cases we find that the anomaly (i.e. the gauge variation of the $d$-dimensional action) is a function of the $G$ gauge field only. Then  the anomaly can be canceled by the anomaly inflow from a $(d+1)$-dimensional topological gauge theory with gauge group $G$. These gauge theories are theories of Dijkgraaf-Witten type and are classified by elements of $H^{d+1}(BG,U(1))$. From the condensed matter perspective, this means that such a $d$-dimensional theory can be realized on the surface of an SPT phase in $d+1$ dimensions with symmetry $G$.

In other cases we find that the anomaly depends on the $d$-dimensional fields and therefore cannot be canceled by the anomaly inflow from a DW theory in $d+1$ dimensions with symmetry $G$. This means that such theories cannot be realized on a surface of an SPT phase. However, it might be possible to realize them on the surface of a Symmetry Enhanced Topological (SET) phase. 

We neglect gravitational and mixed gauge-gravitational anomalies throughout. It has been proposed \cite{Kapustin} that they can be incorporated by replacing $H^{d+1}(BG,U(1))$ with the cobordism group of $BG$  with $U(1)$ coefficients.

It is interesting to compare 't Hooft  anomalies in theories of weakly-interacting fermions and in bosonic gauge theories. We emphasize that the former class of theories does not manifest the most general possible type of anomalies. First of all, for these theories anomalies vanish in odd space-time dimensions. Second, when they are nonzero, they can always be canceled by anomaly inflow and therefore can be characterized by an element of $H^{d+1}(BG,U(1))$. Third, in even space-time dimension $d$ not every element of $H^{d+1}(BG,U(1))$ can arise as an 't Hooft anomaly in a theory of weakly-interacting fermions. For example, we will see that for abelian $G$ the most general anomaly for $d=2$ is cubic in the gauge fields, while the free fermion anomaly is always quadratic. For $d=4$ the most general anomaly is quintic in the gauge fields, while the free fermion anomaly comes from the triangle graph and therefore is cubic. 

{\it Note added.} This paper is an extended version of \cite{KTshort}. While we were writing it up, several other papers with overlapping results have appeared \cite{Ryuetal, Wenetalanomalies, Chenetal}. In particular, it was noted in Ref. \cite{Wenetalanomalies} that in $d=2$ free fermion theories can only give quadratic anomalies, while the most general anomaly which can be canceled by anomaly inflow is cubic. Ref. \cite{Chenetal} noted that there exist $d=3$ theories with 't Hooft anomalies that cannot be canceled  by anomaly inflow from a 4d DW theory. Our results agree with these papers, but the methods are rather different. In particular, the field-theoretic approach we develop works in  all space-time dimensions.

\section{Generalities on anomalies}

Let $G$ be a global symmetry of a $d$-dimensional theory. We would like to promote $G$ to a gauge symmetry and to couple the theory to a gauge field $A$ for $G$ in a gauge-invariant way. There might be obstructions to doing this when the coupling of the theory to $A$ is such that the action cannot be made gauge-invariant. 

A good starting point for understanding these obstructions is the anomaly-inflow assumption, which posits that a theory in $d$ space-time dimensions with an 't Hooft anomaly can be placed on the boundary of topological gauge theory  in $d+1$ dimensions so that the composite system is anomaly-free. It is assumed that the only field in the $(d+1)$-dimensional gauge theory is $A$. It is implicit in this assumption that the gauge variation of the $d$-dimensional action depends only on $A$. Understanding the anomaly then amounts to identifying and studying the $(d+1)$-dimensional TQFT. 

Topological gauge theories are well understood. If we focus on the case of a finite symmetry group $G$, such theories are called Dijkgraaf-Witten theories, and their actions are classified by elements of  $H^{d+1}(BG,U(1))$, the cohomology group of the classifying space of $G$-bundles \cite{DW}. We will say that an anomaly is of DW type if it can be canceled by coupling the theory to a $(d+1)$-dimensional DW theory with gauge group $G$. A natural question is:
\begin{q}
What conditions on a theory and the action of $G$ on it ensure that the 't Hooft anomaly is of DW type?
\end{q}
A related question is
\begin{q}
How do we calculate the class of the DW anomaly theory in these cases?
\end{q}

We address these problems by considering examples in 2d, 3d, and 4d. The boundary theories we consider are themselves topological gauge theories, perhaps coupled to matter. Specifically, in the examples we consider the $d$-dimensional gauge group is either $U(1)$ or a finite abelian group $D$. 

In the latter case we can provide a fairly complete answer to question 1 . The $d$-dimensional theory has a topological action for a $D$-gauge field $a$ specified by a class $\omega \in H^d(BD,U(1))$. A $G$ symmetry of such a theory amounts to an action of $G$ on $BD$ preserving the class $\omega$. The gauge field of the $d$-dimensional theory is a map from the space-time $X$ to $BD$, and the action of $G$ on this field is via the action on $BD$. Note that $G$ acts on the gauge field itself, not simply its gauge equivalence classes. Thus, $G$ in a certain sense can act projectively. This is a potential source of the 't Hooft anomaly \cite{VS}.

The action of $G$ on $BD$ is equivalently an action of $G$ on $D$ together with a class $c \in H^2(BG,D)$, the group cohomology twisted by this action. The action of $G$ on $D$ can be thought of as $G$ permuting the fluxes of the boundary theory. The class $c$ measures the projectivity of this action. 

More formally, the class $c$ defines an extension of groups
$$
1 \to D \to \hat G \to G \to 1.
$$
When we gauge the $G$ symmetry, the original $D$ gauge field and the new $G$ gauge field combine to form a $\hat G$ gauge field. Thus, the gauged theory has a single field, which can be thought of as a map to $B\hat G$.

Since $\hG$ is finite, an action should be specified by a class $\hat \omega \in H^d(B \hat G, U(1))$. The map $D \to \hat G$ defines an embedding
$$
BD \to B \hat G.
$$
The class $\hat \omega$ must restrict to $\omega$ under this inclusion. For a fixed $\omega$, it is not always possible to find such a class $\hat \omega$. These are the cases in which we say there is an 't Hooft anomaly.

How do we go from this description of an anomaly to the $(d+1)$-dimensional anomaly theory? The obstructions to extending $\omega$ are packaged in the Lyndon-Hochschild-Serre spectral sequence of the group extension above \cite{Evens}. These obstructions come one by one as cohomology classes in
$$
H^2(BG, H^{d-1}(BD,U(1))), H^3(BG, H^{d-2}(BD, U(1))), \ldots , H^{d+1}(BG,U(1)).
$$
The vanishing of each class allows the definition of the next obstruction. These are the differentials in the spectral sequence.

Thus, in order for the anomaly to be of DW type, all but the last obstruction must vanish. In this case mathematicians say $\omega$ is transgressive. There are conditions for classes to be trangressive which are well-known in the mathematical literature, answering question 1 for these examples.

Below, we calculate by hand these obstructions in 2d, 3d, and 4d examples of this sort, answering question 2 for these examples. We also perform similar computations for $U(1)$ Chern-Simons theory in $d=3$.  In all these computations, the matter fields play no role, except that they provide a mechanism to choose the extension class $c$. Some examples of this have been explained in \cite{KTshort} and will be recalled below.

The fact that not all classes $\omega$ are transgressive leads to another interesting question:
\begin{q}
What are the anomaly theories in the cases when the lower obstructions are non-vanishing?
\end{q}
Conversely, we can ask:
\begin{q}
Given an anomaly theory, especially a DW theory, how does one produce examples of boundary theories which have the specified anomaly?
\end{q}
We will comment on this question throughout, but we give no systematic method.

\section{Anomalies in 2d bosonic theories}

\subsection{Anomalies in 2d Dijkgraaf-Witten theories}

We begin our study with 't Hooft anomalies in 2d topological gauge theories (perhaps coupled to matter). Previously, 2d 't Hooft anomalies have been discussed in the context of asymmetric orbifolds \cite{Vafa,FreedVafa}. Accordingly, the theories involved scalars as well as free fermions, and discrete symmetries were realized as transformations of the target space which also act on the fermions . The theories we consider are different, as they involve 2d gauge fields. Nevertheless, we will see that some of the anomalies have a similar nature.

We assume that the global symmetry group $G$ is extended by a finite abelian gauge group $D$. The group $G$ may act nontrivially on $D$, so we do not assume in this section that the extension is central. Extensions of $G$ by $D$ are labeled by elements of the abelian group $H^2(BG,D)$ (if $G$ acts nontrivially on $D$, this is group cohomology with twisted coefficients). As explained in the previous section, we expect that the leading obstruction to gauging $G$ lies in in $H^2(BG,H^1(BD,U(1))=H^2(BG,D^*)$, where $D^*=\Hom(D,U(1))$ is the Pontryagin dual of $D$. If this obstruction vanishes, then there can be another obstruction lying in $H^3(BG,U(1))$. If it vanishes too, the symmetry $G$ can be gauged. If it does not vanish, then the 't Hooft anomaly can be canceled by coupling the 2d gauge theory to a 3d DW gauge theory for $G$.

Our goal is to compute both the leading and the subleading anomaly. For simplicity, we consider 2d DW theories with gauge group 
$D=\ZZ_n^N$. Such theories are classified by elements of $H^2(BD, U(1))$. By the universal coefficient theorem, this is $\Hom(H_2(BD),U(1))$. By the K\"unneth formula, 
$$
H_2(B(G_1\times\ldots \times G_n)) \simeq \oplus_{i<j} H_1(BG_i) \otimes H_1(BG_j)
$$
Combining these two, we get that
$$
H^2(B\ZZ_n^N,U(1))\simeq \ZZ_n^{N(N-1)/2}
$$
Thus, possible DW actions can be parameterized by an integral skew-symmetric matrix $\omega_{ij}$ whose elements are defined modulo $n$.

To write a concrete formula for the action, we will use a lattice formulation\footnote{A continuum formulation also exists and is explained in \cite{KS}.} of the DW theory \cite{DW}. In this formulation, one chooses a triangulation $K$ of the 2d space-time $X$ and represents the $\ZZ_n^N$ gauge field by a simplicial 1-cocycle $a$ with values in $\ZZ_n^N$. The path-integral becomes a sum over all such cocycles, with the weight written as $\exp(2\pi S_{2d})$  where
$$
S_{2d} =\frac{k}{n}\int_X \sum_{i<j} \omega_{ij} a^i\cup a^j.
$$
To write the action, we also assumed that $\ZZ_n$ gauge fields are represented by $\ZZ$-valued 1-cochains $a^i$, $i=1,\ldots,N$, which are closed modulo $n$. It is obvious that $\exp(2\pi S_{2d})$ depends only on the value of $a_i$ modulo $n$.  

Let $G$ be a finite global symmetry group for this theory. In general, it may acts nontrivially on the gauge fields $a^i$; this action makes $D$ into a $G$-module. Since $G$ is a symmetry, the pairing $\omega$ must be $G$-invariant. To simplify the matters further, we will assume that $G$ does not mix $a^i$ and $a^j$ for $i\neq j$, i.e. it acts on each $\ZZ_n$ factor separately.

We also need to specify the class of the extension
$$
D\ra \hat G\ra G,
$$
where $\hat G$ is the total symmetry. A nontrivial extension class can be forced on us if an action of $G$ on matter fields closes only modulo elements of the gauge group $D$. Alternatively, we may regard the extension class as part of the definition of the action of $G$ on the 2d DW theory. 

The extension class $c$ takes values in $H^2(BG,D)$, where the action of $G$ on $D=\ZZ_n^N$ can be nontrivial. We can write $c=(c^1,\ldots,c^N)$, where each $c^i$ is a twisted $\ZZ_n$-valued 2-cocycle on $BG$. 

Gauging $G$ means coupling the theory to a $G$ gauge field $A$.  This means, first of all, that we must modify the constraint on $a$ as follows:
$$
\delta_A a = c(A),
$$
where $\delta_A: C^p(X,D)\ra C^{p+1}(X,D)$ is the simplicial differential twisted by the action of $G$ on $D$, and $c(A)\in C^2(X,D)$ denotes the twisted 2-cocycle on $X$ which is the pull-back
of $c\in C^2(BG,D)$ by  the map $\mathcal A:X\ra BG$ corresponding to the gauge field $A$.

The most general ansatz for the gauged action is
\begin{equation}\label{stwodprime}
S_{2d}' = \int_X \hat \omega = \frac{1}{n} \int_X\left( k\sum_{i<j} \omega_{ij} a^i \cup a^j +  \sum_i \left(a^i \cup \Gamma_i(A) + B_i \cup_1 a^i\right) + n \Omega(A)\right),
\end{equation}
where $\Gamma_i(A)$, $B_i(A)$ and $\Omega(A)$ are pull-backs of $\Gamma\in C^1(BG,D^*)$, $B\in C^2(BG,D^*)$ and $\Omega\in C^2(BG,\RR/\ZZ)$  by the map $\mathcal A:X\ra BG$ corresponding to the gauge field $A$. It is understood here that $D^*$ (the Pontryagin dual of $D$) is acted upon by $G$, and this action is dual to the action of $G$ on $D$, so that the pairing $D\times D^*\ra\RR/\ZZ$ is $G$-invariant. This makes the above action invariant under constant $G$ transformations. Note also that we allowed for a coupling involving a higher cup product of Steenrod \cite{Steenrod}. This term is needed on the lattice to correct for the failure of the cup product to be supercommutative on the cochain level. See Appendix B.1 of \cite{KS} for the definition and properties of $\cup_1$. 

For $k=0$ the 1-cochain $\Gamma\in C^1(BG,D^*)$ can be thought of as a magnetic analog of  $c\in C^2(BG,D)$ in the following sense. We can impose the condition
$$
\delta_A a = c(A)
$$
via a Lagrange multiplier $\phi$ which is a 0-cochain on $K$ with values in $D^*\simeq \ZZ_n^N$. The action is then
\begin{multline}
S_{2d}' = \int_X \hat \omega =\frac{1}{n} \int_X\left( k\sum_{i<j} \omega_{ij} a^i \cup a^j +\phi^i\cup  \delta a_i+\right.\\
\left.  +\sum_i \left(-\phi_i \cup c^i+a^i \cup \Gamma_i(A) + B_i \cup_1 a^i\right) + n \Omega(A)\right).
\end{multline}
For $k=0$ there is an obvious parallel between $a,c$ on one hand and $\phi,\Gamma$ on the other hand. In fact, $\exp(2\pi i \phi_i/n)$ can be thought of as an operator which inserts a vortex for $a^i$, so $\phi$ is the magnetic dual of $a$. The equation of motion for $a$ reads
$$
\delta\phi_i+\Gamma_i(A)=0,
$$
which means that the lattice fields $\phi_i$ must transform nontrivially under $G$, in a manner determined by $\Gamma$.

The 't Hooft anomaly is an obstruction to finding a 2d action which is $\hG$-invariant. As explained in \cite{DW}, gauge-invariance of the action is equivalent to the following condition. We imagine that $K$ is a connected component of  the boundary of some three-dimensional CW-complex $J$ and that the gauge fields $(a,A)$ extend to $J$ so that the condition $\delta a=c(A)$ is maintained. The integrand in (\ref{stwodprime}) can be regarded as a 2-cochain $\hat\omega$ on $J$ with values in $\RR/\ZZ$, and the gauge-invariance condition is that this 2-cochain is closed for any $(a,A)$ satisfying the constraint.

A short computation gives
\begin{multline}
\delta\hat\omega=-\frac{k}{n}\sum_{i,j}\omega_{ij} a^i c^j+\frac{k}{n} \sum_{i<j} \omega_{ij} \left(\delta(c^i\cup_1 a^j)-c^i\cup c^j\right)+\\
+\frac{1}{n}\sum_i \left(- a^i\cup\delta_A\Gamma_i+c^i\cup\Gamma_i+\delta(B_i \cup_1 a^i)\right)+\delta\Omega.
\end{multline}

To cancel terms in $\delta\hat\omega$ which are linear in $a$, we need to have (modulo $n$):
$$
B_i=k \sum_{j<i} \omega_{ij} c^j,\quad \delta\Gamma_i=-k \sum_j \omega_{ij} c^j.
$$
The second of these conditions means that the twisted 1-cochain with values in $D^*\simeq\ZZ_n^N$ and components
$$
k \sum_j \omega_{ij} c^j
$$
must be exact. The cohomology class of this 1-cochain, which takes values in $H^1(BG,D^*)$, thus provides an obstruction to gauging $G$. If this obstruction is trivial, then one can ask whether the terms in $\delta\hat\omega$ which are independent of $a$ can be made to vanish as well. This imposes the following constraint:
$$
\delta\Omega=-\frac{1}{n}\sum_i c^i\cup \Gamma_i+\frac{k}{n} \sum_{i<j} \omega_{ij} c^i\cup_1 c^j
$$
It is easy to check that the right-hand side of this equation is closed and thus defines an element of $H^3(BG,\RR/\ZZ)$.The second and last obstruction to gauging $G$ is the vanishing of this element. If it does not vanish, then $G$ cannot be gauged in two dimensions, but the anomaly can be canceled by a 3d DW theory with gauge group $G$.

\subsection{Examples}

Let us list a few examples of 2d DW theories with gauge group $\ZZ_n \times \ZZ_n$ and non-vanishing 't Hooft anomaly. In the case $k=1$, only the trivial extension
$$
D \to D \rtimes G \to G
$$
can be anomaly-free, so any projective $G$-action causes an anomaly. A simple anomalous example with a trivial action of $G$ on $D$ corresponds to a central extension of $G=\ZZ_n$ by $D=\ZZ_n\times\ZZ_n$:
$$
\ZZ_n \times \ZZ_n \to \ZZ_{n^2} \times \ZZ_n \to \ZZ_n.
$$
To get such a central extension, it is sufficient to have a matter field charged under the first $\ZZ_n$ factor in the gauge group, and require that it had fractional charge $1/n$ under the global symmetry $G\simeq\ZZ_n$. This anomaly cannot be cancelled by a 3d DW theory with gauge group $G$, because the leading obstruction does not vanish. 

On the other hand, if we take $k=0$, then the leading obstruction vanishes regardless of the choice of $c$, and the twisted cochain $\Gamma\in C^1(BG,D^*)$ must be closed:
$$
\delta_A \Gamma_j = 0. 
$$
As discussed above, the choice of 1-cocycles $\Gamma_j$ is still physically relevant (it describes the transformation properties of vortex operators under $G$), and we may choose them to be nontrivial. Consider $G = \ZZ_n$ with the same nontrivial abelian extension as above. We may choose $\Gamma_1(A) = A$, $\Gamma_2 = 0$. The extension cocycle is
$$
c_1(A) = \frac{1}{n} \delta A,\quad c_2=0,
$$
Here we again think of the $\ZZ_n$ gauge field $A$ as an integral 1-cochain which is closed modulo $n$, so $\frac{1}{n}\delta A$ is a well-defined integral 2-cocycle. It is easy to see that the cohomology class of this 2-cocycle depends only on the cohomology class of $A$ modulo $n$.\footnote{This map from $H^1(K,\ZZ_n)$ to $H^2(K,\ZZ)$ is known as the Bockstein homomorphism.}  Then the second obstruction is
$$
-\frac{1}{n^2} A \cup \delta A.
$$
The cohomology class of this 3-cocycle is a pull-back of the generator of $H^3(B\ZZ_n,U(1))\simeq\ZZ_n$. Thus there is an 't Hooft anomaly in this 2d theory which can be cancelled by the basic 3d DW theory with gauge group $\ZZ_n$. From the physical viewpoint, the anomaly is caused by the fact that the vortex for $a_1$ transforms nontrivially  (carries charge $-1$) under the global $\ZZ_n$ symmetry. This follows from the equation of motion $\delta \phi_1=-A$. 

Our final example will have a cubic 't Hooft anomaly. As mentioned in the introduction, this sort of 't Hooft anomaly does not occur in free fermion theories. We take $G = \ZZ_n^3$, $D=\ZZ_n\times\ZZ_n$ and $k=0$.  Since $k=0$, the leading obstruction vanishes. To get a nontrivial second obstruction, we take $\Gamma_1(A) = A_3$, $\Gamma_2 = 0$, $c_1(A) = A_1\cup A_2$, and $c_2 = 0$. The extension defined by this $c$ looks as follows:
$$
\ZZ_n^2 \to \cH_n \times \ZZ_n^2 \to \ZZ_n^3.
$$
Here $\cH_n$ is the discrete Heisenberg group of order $n^3$ which is a central extension of $\ZZ_n^2$ by $\ZZ_n$. Such an extension is forced on us, for example, if we have $n$ matter fields on which the generators of $G$ act as clock and shift matrices $x,y$ satisfying
$$
x^n=1,\quad y^n=1,\quad yx=\eta xy,
$$
where $\eta=\exp(2\pi i/n)$.
The second obstruction is non-vanishing and is given by
$$
-\frac{1}{n} A_1\cup A_2\cup A_3.
$$
This 't Hooft anomaly can be canceled by coupling to a DW theory with gauge group $G=\ZZ_n^3$.

\section{Anomalies in 3d bosonic theories}

\subsection{Anomalies in a 3d $\ZZ_n$ gauge theory}

In this section we study 't Hooft anomalies in 3d topological gauge theories (perhaps coupled to matter). 
One natural class of such theories is 3d DW theories. Another one is 3d Chern-Simons theories. The two classes of theories overlap, but neither is a subset of the other. Nonabelian DW theories are not equivalent to Chern-Simons theories, in general. On the other hand, generic Chern-Simons theories have framing anomalies and therefore do not admit a lattice description, while DW theories are defined on the lattice and therefore are free from framing anomalies. 

Let us first construct an example of a 3d DW theory with an 't Hooft anomaly for $G$.  We assume for simplicity that the 3d gauge group $D$ is isomorphic to $\ZZ_n$ and that the global symmetry $G$ acts trivially on $D$. Following the idea of \cite{VS}, we assume that total symmetry group $\hG$ is an extension of $G$ by $D$. Since $G$ acts trivially on $D$ and $D$ is abelian, this is a central extension. Such extensions are classified by elements of $H^2(BG,D)$. Let $c\in H^2(BG,D)$ be the extension class, and $\tc\in C^2(BG,\ZZ)$ be its lift to an integral cochain on $B$. The cochain $\tc$ is closed modulo $n$:
$$
\delta \tc=n \gamma_c,\quad \gamma_c\in C^3(BG,\ZZ). 
$$
Clearly, $\gamma_c$ is an integral 3-cocycle, $\delta\gamma_c=0$. The cohomology class of $\gamma_c$ is well-defined (does not depend on the way one lifts $c$ to $\tc$).  The map $H^2(BG,\ZZ_n)\ra H^3(BG,\ZZ)$ which sends $c$ to $\gamma_c$ is known as the Bockstein homomorphism. In what follows the reduction of $\gamma_c$ modulo $n$ will play an important role; by an abuse of notation we will denote it $\gamma_c$ as well.

Since $H^3(B\ZZ_n,U(1))=\ZZ_n$, 3d DW actions with gauge group $\ZZ_n$ are labeled by an integer $k$ modulo $n$. To write an action explicitly, we assume that our 3-manifold  is equipped with a triangulation $K$. Then a $\ZZ_n$ gauge field can be described by an integral cochain $a\in C^1(K,\ZZ)$ with a constraint $\delta a= n\beta$, where $\beta$ is an integral 2-cochain. The DW action has the form reminiscent of the Chern-Simons action:
$$
S_{3d}=\frac{k}{n^2} \int_K a\cup \delta a=\frac{k}{n}\int_K a\cup\beta.
$$
It is invariant under gauge transformations $a\mapsto a+\delta f$, $f\in C^0(K,\ZZ)$ as well as $a\mapsto a+n\alpha$, $\alpha\in C^1(K,\ZZ)$. The latter gauge symmetry means that only the values of $a$ modulo $n$ are physical.

Once a $G$ gauge field $A$ is introduced, the constraint on $a$ is modified to $\delta a=n\beta+\tc(A)$, where $\tc(A)$  is the pull-back of $\tc\in C^2(BG,\ZZ)$ by the $G$ gauge field $A$ regarded as a simplicial map $A: K\ra BG$. Note that this equation implies $\delta\beta=-\gamma_c(A)$. 

Apart from modifying the constraint, gauging $G$ may also introduce explicit dependence on $A$ into the action. The most general ansatz for the gauged action is
\begin{equation}\label{sprite}
S'_{3d}=\int_K \left(\frac{k}{n} a\cup \beta+B(A)\cup a+a\cup_1 H(A) +\beta\cup_1 B'(A)+\beta\cup_2 H'(A)+\Omega(A)\right),
\end{equation}
where $B(A), B'(A),$ $H(A), H'(A),$ and $\Omega(A)$ are pull-backs of cochains $B,B'\in C^2(BG,\RR/\ZZ)$, $H,H'\in C^3(BG,\RR/\ZZ)$, and $\Omega\in C^3(BG,\RR/\ZZ)$. 

Note that we allowed for couplings involving the higher cup products $\cup_i$ of Steenrod \cite{Steenrod}. These products control the failure of supercommutativity for the ordinary cup product of cochains, see for example appendix B.1 of \cite{KS} for a review. We will see below that these couplings are not independent, i.e. $B',H,H'$ are fixed by other data.

The 2-cochain $B$ plays a role analogous to $\tc$ and can be thought of as the ``magnetic" analogue of $\tc$. This is most obvious in the case $k=0$. The constraint $\delta a=n\beta+\tc$ can be enforced using a Lagrange multiplier $b\in C^1(K,\ZZ)$, and the action including the Lagrange multiplier (and with $k$ set to zero) is
$$
S'_{3d}=\frac{1}{n} \int_K  \left(b\cup \delta a- b\cup \tc(A)+ n B(A)\cup a+\ldots\right),
$$
where dots denote terms involving higher cup products.
This action is symmetric under the exchange of $a$ and $b$ as well as $\tc$ and $-nB$ (and if we also neglect the potential effect of higher cup products). This symmetry exchanges electric and magnetic excitations in the DW theory. Recall that a DW theory has both electric and magnetic quasi-particles. Electric quasi-particles are represented by Wilson loops for $a$, while magnetic quasi-particles are vortices, i.e. modifications of the constraint $\delta a=n\beta$ along a loop on the dual cell complex. When the constraint $\delta a=n\beta$ is enforced using a Lagrange multiplier $b$, magnetic quasi-particles are represented by Wilson loops for $b$. Thus exchanging $a$ and $b$ exchanges electric and magnetic quasi-particles.

An attentive reader might have noticed that the cochain $B$ takes values in $\RR/\ZZ$, while the cochain $\tc$ takes values in $\ZZ$ (or rather $\ZZ_n$, since only the value of $\tc$ modulo $n$ matters). However, in order to ensure the invariance of the action under $a\mapsto a+n\alpha$, $B$ must have the form $B_0/n$, where $B_0$ is an integral cochain. So the electric-magnetic symmetry exchanges the integral 2-cochains $\tc$ and $-B_0$. 

For $k\neq 0$ the symmetry between electric and magnetic quasi-particles is destroyed, since vortices now carry electric charge.
Nevertheless, $\tc$ and $B$ still play symmetric roles, in the sense that $\tc$ and $B$ determine the transformation properties of the two kinds of excitations under $G$ gauge transformations. Therefore we will regard 2-cochains $\tc$ and $B$ as fixed and will try to find the cochains $B',H,H', \Omega$ such that the action is a well-defined action for a DW theory with gauge group $\hG$. Actually, we will find that for $k \neq 0$ there are  certain  compatibility conditions between $\tc$ and $B$ as well, so that the distinct allowed values of $B$ are parameterized by elements of $H^2(BG,\ZZ_n)$, just like $c$. 

The first requirement is the invariance of the action under $a\mapsto a+n\alpha$ and $\beta\mapsto \beta+\delta \alpha$, where $\alpha\in C^1(K,\ZZ)$ is arbitrary. This invariance ensures that only values of $a$ modulo $n$ have a physical meaning. While for $\tc=0$ the first term in action is invariant, for $\tc\neq 0$ is transforms as follows:
$$
\int_K \frac{k}{n} a\cup\beta\mapsto \int_K \frac{k}{n} a\cup\beta+\int_K \frac{k}{n} \tc\cup \alpha.
$$ 
To ensure that the action is invariant, we need to take
$$
B=-\frac{k}{n^2}\tc+\frac{1}{n}B_0,\quad H=\frac{1}{n} H_0, \quad B'=0,\quad H'=0.
$$
where $B_0\in C^2(BG,\ZZ)$, $H_0\in C^3(BG,\ZZ)$. 

The second requirement is gauge-invariance under $\hG$ gauge transformations.
As explained in \cite{DW}, this is equivalent to the following condition. We imagine that $K$ is a connected component of  the boundary of some four-dimensional CW-complex $J$ and that the gauge fields $(a,A)$ extend to $J$ so that the condition $\delta a=n\beta+\tc(A)$ maintained. The integrand in (\ref{sprite}) can be regarded as a 3-cochain $\omega$ on $J$ with values in $\RR/\ZZ$, and the condition is that this 3-form is closed. 

A short computation gives
$$
\delta\omega=-\frac{2k}{n} \gamma_c \cup a+\frac{1}{n}\delta B_0\cup a+\frac{1}{n} \delta \left(a\cup_1\left(-k \gamma_c+H_0\right)\right)+\ldots,
$$
where dots denote terms independent of $a$. This fixes $H_0$:
$$
H_0=k\gamma_c,
$$
and also gives a constraint on $B_0$:
$$
\delta B_0=2k \gamma_c\ {\rm mod}\ n.
$$
That is, the cohomology class of $2k\gamma_c\in H^3(BG,\ZZ_n)$ must be trivial, and $B_0\ {\rm mod}\ n$ should be a trivialization of the corresponding cocycle. For a fixed $[c]\in H^2(BG,\ZZ_n)$ the freedom in the choice of $B_0$ is parameterized by elements of $H^2(BG,\ZZ_n)$. Thus we retain a certain symmetry between $c$ and $B$ even for nonzero $k$. 

The cohomology class $[2k\gamma_c]\in H^3(BG,D^*)$ is the leading obstruction to gauging $G$ in this theory. Note that for more general gauge groups $D$ we also expect obstructions living in $H^2(BG,H^2(BD,U(1)))$. In our case this obstruction vanishes because $H^2(B\ZZ_n,U(1))=0$. 

Assuming that the leading obstruction vanishes, let us look at the remaining $a$-independent terms in $\delta\omega$. They read:
$$
-\frac{k}{n^2} \tc\cup\tc+\frac{k}{n} \tc\cup_1\gamma_c+\frac{1}{n}B_0\cup\tc+\delta\Omega. 
$$
For this to vanish, the cohomology class of the 4-cocycle 
$$
\frP_k(c,B_0)=-\frac{k}{n^2} \tc\cup\tc+\frac{k}{n} \tc\cup_1\gamma_c+\frac{1}{n}B_0\cup\tc
$$
should be trivial.
One can easily check that $\frP_k(c,B_0)\in C^4(BG,\RR/\ZZ)$ is closed. One can also check that the cohomology class of $\frP_k(c,B_0)$ depends only on the cohomology class of $c$ and the choice of $B_0$ modulo $n$, but does not depend on the choice of $\tc$ and does not change if one shifts $B_0$ by an exact 2-cocycle. 

The cohomology class of $\frP_k(c,B_0)$ is the second (and last) obstruction for gauging $G$, and it takes values in $H^4(BG,\RR/\ZZ)$, as expected. It simplifies in special cases. First, consider the case $k=0$. Then the constraint on $\gamma_c$ becomes vacuous, and the constraint on $B_0$ becomes $\delta B_0=0\, {\rm mod}\, n$, i.e. it is a 2-cocycle on $BG$ modulo $n$, just like $c$. The cohomology class of the 4-cocycle $\frP_k(c,B_0)$ simplifies to
$$
\iota_n([B_0\cup c]),
$$
where here and below we denote by $\iota_m$ the obvious embedding $H^4(BG,\ZZ_m)\subset H^4(BG,U(1))$. Thus for $k=0$ the anomaly arises solely from the ``interference'' between $c$ and $B_0$. In particular, for a fixed extension class $[c]$ one can always avoid an anomaly by setting $B_0=0$.

Another fairly simple case is $n=2$. In that case the only nonzero value of $k$ is $k=1$. The constraint on $\gamma_c$ is again vacuous, and $B_0$ is again a 2-cocycle. The  cohomology class of the 4-cocycle $\frP(c,B_0)$ can be written as
$$
\iota_4(\frP(c))+\iota_2([B_0\cup c]),
$$
where $\frP(c)\in H^4(BG,\ZZ_4)$ is the Pontryagin square of $c$ (see e.g. the appendix in \cite{KapThoone} for a definition of the Pontryagin square). Thus for $n=2$ and $k=1$ the anomaly can be nonzero even for vanishing $B_0$.

\subsection{Examples}

Let us list a few examples of 3d DW theories with gauge group $\ZZ_n$  where the 't Hooft anomaly is non-vanishing. Let $n=2$, $k=0$ and $G=\ZZ_2\times\ZZ_2$.  In this case $H^*(G,\ZZ_n)$ is an algebra of polynomials in two variables $x,y$ of degree $1$ with coefficients in $\FF_2$ (the field with two elements) \cite{Evens}. Thus $H^2(BG,\ZZ_2)$ is a 3-dimensional vector space over $\FF_2$ with a basis $x^2,y^2,xy$. The extensions not involving $xy$ are abelian, they correspond to $\hG\simeq \ZZ_4\times\ZZ_2$. The extensions involving $xy$ are nonabelian. Specifically, if the extension class is $c=xy+x^2+y^2$, the group $\hG$ is isomorphic to the group of quaternionic units $Q_8$. This is a finite subgroup of $SU(2)$ generated by $i\sigma_1,$ $i\sigma_2$ and $i\sigma_3$. The extension classes $c=xy$, $c=x^2+xy$ and $c=y^2+xy$ correspond to $\hG$ isomorphic to the dihedral group of order $8$ which we denote $D_8$. Irreducible representations of $Q_8$ and $D_8$ which are nontrivial under $\ZZ_2\times\ZZ_2$ are two-dimensional.  

To get an 't Hooft anomaly, we need to choose two extension classes $[c]$ and $[B_0]$  in $H^2(BG,\ZZ_2)$ which control the transformation properties of electric and magnetic excitations so that $c\cup B_0$ is nonzero when mapped to $H^4(BG,U(1))$. It can be shown that the only elements in
$H^4(BG,\ZZ_2)$ which remain nonzero under this map are $xy^3$ and $y x^3$ \cite{Drinfeld}. This implies, first of all, that if both $c$ and $B_0$ correspond to an abelian extension, the 't Hooft anomaly vanishes. If one of the extension classes is abelian, and the other one is nonabelian, the 't Hooft anomaly is nonzero. Finally, if both $c$ and $B_0$ correspond to nonabelian extensions, the 't Hooft anomaly vanishes if and only if $c=B_0$. Thus to get a non-vanishing 't Hooft anomaly it is sufficient to have electric excitations transforming in a two-dimensional irreducible representations of either $D_8$ or $Q_8$, and to have magnetic excitations transforming in a different projective representation of $G=\ZZ_2\times\ZZ_2$.

Another class of examples is obtained by letting $n=p$, an odd prime and $G=\ZZ_p\times\ZZ_p$. In this case $H^*(D,\ZZ_p)$ is a supercommutative algebra over $\FF_p$ (the field with $p$ elements) with two odd generators $x,y$ in degree $1$ and two even generators $\beta_x$ and $\beta_y$ in degree $2$ \cite{Evens} ($\beta_x$ and $\beta_y$ are the image of $x$ and $y$ under the Bockstein homomorphism $H^1(G,\ZZ_p)\ra H^2(G,\ZZ_p)$). Thus $H^2(G,D)$ is a three-dimensional vector space over $\FF_p$ with a basis $xy$, $\beta_x$ and $\beta_y$. Linear combinations of $\beta_x$ and $\beta_y$ correspond to abelian extensions, while any extension class which contains $xy$ corresponds to a nonabelian extension. If the extension is nonabelian, the group $\hG$ has an irreducible $p$-dimensional representation.

The Bockstein homomorphism $H^2(G,\ZZ_p)\ra H^3(G,\ZZ_p)$ maps $\beta_x$ and $\beta_y$ to zero and maps $xy$ to $\beta_x y+x \beta_y$. Thus to get a theory with a nonvanishing 't Hooft anomaly it is sufficient to take $k\neq 0$ modulo $p$ and let $c$ be any nonabelian extension. Then $\gamma_c\neq 0$, and the symmetry $\ZZ_p\times\ZZ_p$ cannot be gauged, regardless of the choice of $B_0$.  Note that in this case one cannot cancel the anomaly by coupling the 3d theory to a 4d DW theory with gauge group $G$, since the gauge variation of the action depends on the 3d gauge field $a$.

Alternatively, as for $n=2$, we can take $k=0$ and choose $c$ and $B_0$ so that their cup product does not vanish when mapped to $H^4(BG,U(1))$. This requires at least one of the extension classes $c$ and $B_0$ to be nonabelian \cite{Drinfeld}. A slight difference compared to the case $n=2$ is that now one can have $c=B_0$ and a nonzero 't Hooft anomaly. 

\subsection{An aside on Deligne-Beilinson cocycles}

Below we will need to manipulate topological gauge-invariant actions constructed out of $U(1)$ gauge fields. One way to get such an action is to integrate a gauge-invariant form of top degree over the space-time manifold. Such a form is a polynomial built from curvature 2-forms of the gauge fields. However, one often encounters actions which are gauge-invariant only modulo elements of $2\pi\ZZ$, like Chern-Simons actions and their generalizations. Such actions are not integrals of globally-defined forms. To deal with such actions the formalism of Deligne-Beilinson cocycles is very convenient.  In this section we give a brief review of this formalism, see e.g. \cite{DB} for more details. This material can be omitted at first reading, if one is prepared to take on faith that certain formulas we write in the next subsection and in section \ref{sec:fourd} are well-defined.

Let $A$ be a smooth $q$-form, $q>0$. A $(q-1)$-form gauge transformation of $A$ is a transformation $A\mapsto A+d\lambda$, where $\lambda$ is a $(q-1)$-form. We can extend this definition to $q=0$ by defining a $(-1)$-form as a locally constant real function with values in $2\pi\ZZ$, and defining a gauge transformation of a 0-form  $f$ by a $(-1)$-form $e$ as a transformation $f\mapsto f+e$.  Thus for $(-1)$-forms  the role of $d$ is played by the embedding $2\pi\ZZ\ra \RR$. Note that this definition maintains the property $d^2=0$. Note also that a $0$-form modulo a $(-1)$-form gauge transformation is a $2\pi$-periodic scalar. 

Informally, a DB cocycle of degree $q$ is defined by a collection of smooth $q$-forms on coordinate charts, which on double overlaps are related by smooth $(q-1)$-form gauge transformations. These $(q-1)$-form gauge transformations are related by $(q-2)$-form gauge transformations on triple overlaps, etc. A DB $q$-cocycle has a $q$-component, $(q-1)$-component, etc., all the way down to a $(-1)$-component. The $q$-component lives on coordinate charts $U_i$, the $(q-1)$-component lives on double overlaps $U_{ij}=U_i\bigcap U_j$, etc. 

More formally, consider the extended de Rham-Cech bi-complex $\Omega^{p,r}$, $p=-1,0,\ldots,$ $q=0,1,\ldots$ with respect to some open cover of $X$. Here $p$ is the de Rham degree and $q$ is the Cech degree. We denote by $d$ and $\delta$ the de Rham and Cech differentials, respectively. A DB cocycle of degree $q$ is an element 
$$
A=\sum_{i=-1}^q A_i\in \bigoplus_{i=-1}^q \Omega^{i,q-i}
$$
satisfying 
$$
\delta A_i=d A_{i-1}, \quad i=-1,\ldots, q.
$$

Let us give a few examples. For simplicity let us assume that all $U_i$ and their multiple overlaps are connected. Then a DB $0$-cocycle is given by a collection of functions $f_i: U_i\ra \RR$ and a collections of numbers $e_{ij}\in 2\pi \ZZ$ on double overlaps $U_{ij}$ such that $f_i-f_j=e_{ij}$. A DB $1$-cocycle is a collection of 1-forms $A_i\in \Omega^1(U_i,\RR)$, smooth functions $f_{ij}:U_{ij}\ra \RR$ and numbers $e_{ijk}\in 2\pi\ZZ$ such that $A_i-A_j=df_{ij}$ and $f_{ij}+f_{jk}+f_{ki}=e_{ijk}$. 

The basic property of a DB $q$-cocycle $A$ is that it can be integrated over a smooth $q$-chain $X$, the result being a real number defined modulo $2\pi \ZZ$. This number will be denoted by $\int_X A$. Applying to this number the function $x\mapsto \exp(i x)$, we get an element of $U(1)$ which can be called the holonomy of a DB $q$-cocycle on the $q$-chain $X$. Another basic operation is the exterior derivative which maps a DB $q$-cocycle $A$ to a DB $q+1$-cocycle $dA$. This operation acts as the usual exterior derivative on components of non-negative degree; for the component of degree $-1$ it acts by means of the embedding $2\pi \ZZ\ra \RR$. The top component of $dA$ is a closed $(q+1)$-form, and it can be integrated over any smooth $(q+1)$-chain $Y$ (not just a $(q+1)$-cycle). If $Y$ is actually a cycle, then the integral takes value in $2\pi\ZZ$. More generally, there is a form of Stokes theorem:
$$
\int_{\partial Y} A=\int _Y dA\ {\rm mod}\ 2\pi\ZZ.
$$
The bottom components of $dA$ are also special: its $(-1)$-component is zero, while its $0$-component consists of functions on $(q+2)$-tuple overlaps which take values in $2\pi\ZZ$.  

The Stokes formula can be taken as the defining property of DB cocycles. In this approach a DB cocycle is known as a Cheeger-Simons differential character. It is more convenient for our purposes to work with the definition using coordinate charts and their overlaps. 

A DB cocycle of the form $dA$, where $A$ is another DB cocycle, is called an exact DB cocycle. An exact DB $q$-cocycle is merely a collection of Cech cocycles with values in locally exact differential forms, with form degree varying from $0$ to $q$ and Cech degree varying from $0$ to $q$, so that the total degree is $q$. It follows from the Stokes formula that exact DB  cocycles have trivial holonomy. Here are a few examples. An exact DB $0$-cocycle is a smooth function which takes values in $2\pi\ZZ$. An exact DB $1$-cocycle is a connection on a $U(1)$ bundle which is a pure gauge. 

The top component of an exact DB $q$-cocycle on $X$ defines an element of $H^q(X,\RR)$. The bottom component defines an element of $H^q(X, 2\pi\ZZ)$. These two elements are not independent: the former is the image of the latter under the embedding $2\pi\ZZ\ra \RR$. In the case $q=1$ this cohomology class encodes the winding numbers of the $2\pi$-periodic scalar. In the case $q=2$ this cohomology class is $2\pi$ times the first Chern class of the connection. 

DB cocycles can be multiplied by integers and added, thus they form a (graded) abelian group. They do not form an algebra (there is no way to multiply two generic DB cocycles). However, exact DB cocycles divided by $2\pi$ can be multiplied in the usual way and form an associative graded algebra.\footnote{This algebra is not supercommutative, unfortunately. This happens because the product of Cech cocycles is not supercommutative in degree greater than $0$. However, it is supercommutative up to $\delta$-exact terms, which drop out upon integration over a closed manifold. This is good enough for our purposes, so we can manipulate the actions as if the product were supercommutative.} Generic DB cocycles form a module over this algebra. That is, one can multiply an arbitrary DB $q$-cocycle by $1/(2\pi)$ times an exact DB $p$-cocycle and get a DB $(p+q)$-cocycle. For $p=0$ this is simply the operation of multiplication of a DB cocycle by an integer.

In general, one cannot divide a DB cocycle by an integer, so it may happen that an integer multiple of a DB cocycle is exact, but the cocycle itself is not exact. For example, if a $2\pi$-periodic scalar $\phi$ takes values in $\frac{1}{n} 2\pi\ZZ$, $n\phi $ is an exact DB $0$-cocycle, but $\phi$ itself is not exact. Similarly, if $A$ is a flat connection on a line bundle with holonomy taking values in $n^{\rm th}$ roots of unity, $nA$ is a pure gauge, i.e. $nA=d\phi$ for some periodic scalar $\phi$, but $A$ itself need not be pure gauge. In what follows we will often make use of such connections. Note that while the top component of the 2-cocycle $dA$ in this case vanishes (because the connection is flat), the cocycle itself need not vanish. The 1st Chern class of such a connection can be nonzero, but multiplying it by $n$ gives zero, i.e. it is $n$-torsion. This follows formally by writing $n dA=d(nA)=d(d\phi)=0$. Note also that the Chern-Simons ``forms'' $A(dA)^p/(2\pi)^p$ are well-defined DB cocycles of degree $2p+1$. These cocycle are not exact, but if $nA$ is exact, multiplying the Chern-Simons ``form'' by $n$ gives an exact DB cocycle, since 
$$
\frac{1}{(2\pi)^p} nA(dA)^p=\frac{1}{(2\pi)^p} d\phi (dA)^p=\frac{1}{(2\pi)^p}
d(\phi (dA)^p).
$$
It follows that for a closed $(2p+1)$-manifold $X$ and a connection $A$ satisfying $nA=d\phi$ the Chern-Simons action 
$$
\frac{1}{(2\pi)^p} \int_X A (dA)^p
$$
is an integer multiple of $2\pi/n$.

\subsection{Anomalies in a 3d Chern-Simons theory}

Let us now consider $U(1)$ Chern-Simons theory at an even level $2k$. The level must be even so that the Chern-Simons action is topological rather than spin-topological \cite{DW}. This theory has a framing anomaly \cite{Witten_Jones} and thus cannot be defined by a lattice action. However, it can be described by a continuum action
$$
S_{3d}=\frac{k}{2\pi}\int_X a da.
$$
Here $a$ is a $U(1)$ gauge field which transforms as $a\mapsto a+df$ under $U(1)$ gauge transformations, with $f$ a $2\pi$-periodic scalar. Equivalently, one can say that $a$ is a DB $1$-cocycle and $f$ is a DB $0$-cocycle, while the action density is a well-defined DB $3$-cocycle.

We do not know how to handle the case of a general finite global symmetry group $G$ in the continuum, so we limit ourselves to the case $G=\ZZ_n\times\ZZ_n$. In that case we can describe $G$ gauge fields by a pair of $U(1)$ gauge fields $A_1$ and $A_2$ and a pair of $2\pi$-periodic St\"uckelberg ghosts $\phi_1$ and $\phi_2$ which transform as follows:
$$
A_i\mapsto A_i+df_i,\quad \phi_i\mapsto \phi_i+nf_i.
$$
Here $f_i$, $i=1,2$ are also $2\pi$-periodic scalars.  We also have constraints $d\phi_i=n_i A_i$. They can be imposed by hand, or by means of Lagrange multiplier fields.

The most general action is
\begin{equation}\label{CS3d}
S'_{3d}=\int_X\left( \frac{k}{2\pi} a da+ \frac{1}{2\pi} a (p_1 dA_1+p_2 dA_2)+ \frac{p_3}{(2\pi)^2} a d\phi_1 d\phi_2\right).
\end{equation}
The coefficients $p_1,p_2$ and $p_3$ must be integral to ensure invariance under $a\mapsto a+df$ (here we make use of the fact that if $\phi$ is a $2\pi$-periodic scalar, then $d\phi$ is a closed 1-form whose periods are integer multiples of $2\pi$). Another way to explain the integrality of $p_i$ is to say that we want the action density to be a well-defined DB $3$-cocycle.

The situation with $\ZZ_n\times\ZZ_n$ gauge symmetry is more complicated. In our description, $\ZZ_n\times\ZZ_n$ arises as a subgroup of $U(1)\times U(1)$ which leaves $\phi_1$ and $\phi_2$ invariant. To ensure local $\ZZ_n\times\ZZ_n$ symmetry we need to find a suitable transformation law for $a$ under $U(1)\times U(1)$ which makes the action gauge-invariant. When $p_3=0$, one can postulate that $a$ does not transform at all under $U(1)\times U(1)$ gauge transformations. In this case there is no 't Hooft anomaly. For $p_3\neq 0$ this no longer works. We can try to cancel at least $a$-dependent terms in the variation of the action.  The transformation law must preserve the quantization condition on the periods of $da$ (they must be integral multiples of $2\pi$). Equivalently, the transformation must respect the fact that $a$ is a DB $1$-cocycle. Modulo 3d $U(1)$ gauge transformations,  the only possible transformation law under $U(1)\times U(1)$ transformation $(f_1,0)$ is
$$
a\mapsto a- q_1 f_1 \frac{d\phi_2}{2\pi},
$$
and under the transformation $(0,f_2)$
$$
a\mapsto a+ q_2 f_2 \frac{d\phi_1}{2\pi}.
$$
Here $q_1$ and $q_2$ are integers.

Requiring the cancelation of $a$-dependent terms in the variation of the action (\ref{CS3d}) we get $q_1=q_2=q$ and
\begin{equation}\label{kprel}
2kq=n p_3.
\end{equation}
This equation need not have integral solutions for an arbitrary choice of $k,n,$ and $p_3$. It is natural to fix $k$ and $n$ and regard (\ref{kprel}) as a constraint on possible choices of $p_3$ and $q$. The general solution is 
$$
q=\frac{n\ell}{gcd(n,2k)},\quad p_3=\frac{2k \ell}{gcd(n,2k)},\quad \ell\in\ZZ.
$$
Note that for $\ell=gcd(n,2k)$ one can define a new gauge field
$$
a'=a+\phi_1 \frac{d\phi_2}{2\pi}
$$ 
which is invariant under $(f_1,0)$ and transforms as follows under $(0,f_2)$:
$$
a'\mapsto a'=a'+\frac{n}{2\pi} d(f_2\phi_1).
$$
This is merely a 3d gauge transformation with a $\phi_1$-dependent parameter. The same redefinition removes the last term in the action (\ref{CS3d}). Thus the theory with $\ell=gcd(n,2k)$ is equivalent to the theory with $\ell=0$. In other words, the transformation 
$$
p_3\mapsto p_3+2k,\quad q\mapsto q+n
$$
is an equivalence of theories. 

Under the transformation $(f_1,0)$ the action (\ref{CS3d}) changes as follows:
$$
S'_{3d}\mapsto S'_{3d}+\frac{q}{(2\pi)^2} \int_X  f_1 d\phi_2 (p_1 dA_1+p_2 dA_2).
$$
Similarly, under the transformation $(0,f_2)$ the action (\ref{CS3d}) changes as follows:
$$
S'_{3d}\mapsto S'_{3d}-\frac{q}{(2\pi)^2} \int_X  f_2 d\phi_1 (p_1 dA_1+p_2 dA_2).
$$
If $q p_1$ and $q p_2$ are integral multiples of $n$,  the 3d action is gauge-invariant, because $ndA_1=ndA_2=0$. Otherwise there is an 't Hooft anomaly. It can be canceled by coupling the 3d theory on $X=\partial Y$ to a topological 4d gauge theory on $Y$ with an action
$$
S_4=-\frac{nq}{(2\pi)^2} \int_Y A_1 A_2 (p_1 dA_1+p_2 dA_2).
$$
This is a continuum description of the 4d DW theory with gauge group $\ZZ_n\times\ZZ_n$. Such actions are parameterized by elements of $H^4(B(\ZZ_n\times\ZZ_n), U(1))=\ZZ_n\times\ZZ_n$. In the above continuum description the parameters are $q p_1$ and $q p_2$ modulo $n$. 

Note that the commutator of the transformations $(f_1,0)$ and $(0,f_2)$ is not zero but is a 3d gauge transformation with a parameter $qn f_1 f_2/2\pi$. This shows that the symmetry of the 3d system is not a product of the gauge $U(1)$ and the global $\ZZ_n\times\ZZ_n$, but an extension of the latter by the former. Such extensions are parameterized by $H^2(B(\ZZ_n\times\ZZ_n),U(1))=\ZZ_n$. In the present case the class of the extension is determined by $q\, {\rm mod}\, n$. The presence of the Chern-Simons coupling at level $2k$ leads to a further constraint on $q$ (namely, $2k q=0\, {\rm mod}\, n$), so that not all extensions classes are realizable in this system. 
 
Finally, let us comment on the physical meaning of the parameters $p_1$ and $p_2$. Wilson lines in Chern-Simons theory may terminate on monopole operators defined by the condition 
$$
\int_{S^2_p} da=2\pi m,
$$
where $S^2_p$ is a small 2-sphere centered at the insertion point. Such an operator has electric charge $2km$, so a Wilson line can terminate on a monopole operator if its charge is divisible by $2k$. (Thus effectively the gauge group becomes $\ZZ_{2k}$). In the presence of a global symmetry $G$ monopole operators may also carry charge under $G$. $p_1$ and $p_2$ are the charges of a unit monopole with respect to $\ZZ_n\times\ZZ_n$. 

\subsection{Examples}

Let $n=2$, $k=1$, $p_3=1$, and $q=1$. In this case the global symmetry group $\ZZ_2\times\ZZ_2$ is realized projectively: the generators of the two $\ZZ_2$ factors anti-commute rather than commute when acting on charge-1 matter fields. For example, one can consider a pair of complex scalar fields, both of charge $1$, on which the two generators act as Pauli matrices $\sigma_1$ and $\sigma_2$. If these fields are also coupled to a $U(1)$ Chern-Simons gauge field at level $2$, the $\ZZ_2\times\ZZ_2$ symmetry acts projectively. Whether or not it can be gauged depends on the parameters $p_1$ and $p_2$ which are defined modulo $2$. These parameters determine the $\ZZ_2\times\ZZ_2$ quantum numbers of monopole operators. If either $p_1$ or $p_2$ are odd, unit monopoles are odd under the generator of at least one of the $\ZZ_2$ factors. In this case the above analysis shows that there is an 't Hooft anomaly which can be canceled by coupling the 3d theory to the 4d DW theory with gauge group $\ZZ_2\times\ZZ_2$.

A different kind of 't Hooft anomaly occurs when the relation (\ref{kprel}) cannot be satisfied. For example, consider $U(1)$ Chern-Simons theory with $k=1$ coupled to three scalar fields with charge $1$. Let us assume that the interactions of the fields are invariant under $U(3)$. The global symmetry group is then $U(3)/U(1)=SU(3)/\ZZ_3$.  It has a subgroup isomorphic to $\ZZ_3\times\ZZ_3$ with generators 
\begin{equation}
x=\begin{pmatrix} 1 & 0 & 0\\  0 &\eta & 0\\ 0 & 0 & \eta^2\end{pmatrix},\ y=\begin{pmatrix} 0 & 1 & 0\\ 0 & 0& 1\\ 1 & 0 & 0\end{pmatrix},\ 
\end{equation}
where  $\eta=\exp(2\pi i/3)$. They satisfy $yx=\eta xy$, i.e. $\ZZ_3\times\ZZ_3$ acts projectively, with the extension parameter $q=1$. But the relation (\ref{kprel}) requires $2=3p_3$, which is impossible to satisfy with an integer $p_3$. Thus $\ZZ_3\times\ZZ_3$ has an 't Hooft anomaly, but it cannot be canceled by coupling to a 4d DW theory with gauge group $\ZZ_3\times\ZZ_3$.
This situation is analogous to the case of cohomologically nontrivial $2k\gamma_c$ in the previous subsection.
 
\section{Anomalies in 4d bosonic theories}\label{sec:fourd}

In 4d, anomalies of connected Lie group symmetries are classified by Chern-Simons actions in 5d. For example, for $G=U(1)$ the Chern-Simons action has the form
$$
S_{5d}=\frac{k}{(2\pi)^2}\int_Y A (dA)^2,
$$
where $k$ is integral. On the boundary of such a 5d theory one should place a 4d theory with an 't Hooft anomaly for a global $U(1)$ symmetry. It could either be a system of free Weyl fermions with charges $Q_i$ satisfying $\sum_i Q_i^3=k$, or a Goldstone boson $\phi$ with an axion coupling
$$
S'_{4d}=\frac{k}{(2\pi)^2}\int_X \phi F\wedge F.
$$

Turning to discrete symmetries, if we assume that the anomaly can be canceled by the anomaly inflow, we need to consider DW actions in 5d which are classified by $H^5(BG,U(1))$. For $G=\ZZ_n$ we have $H^5(BG,U(1))=\ZZ_n$. The corresponding lattice action has the form
$$
S_{5d}=\frac{k}{n^3} \int_Y \hat A(\delta \hat A)^2,
$$
where $\hat A$ is an integral 1-cochain which is closed modulo $n$. It represents the $\ZZ_n$ gauge field. Alternatively, we can write down a continuum action using a $U(1)$ gauge field and a charge-$n$ periodic scalar $\phi$ satisfying the constraint $d\phi=nA$:
$$
S_{5d}=\frac{k}{(2\pi)^2} \int_Y A (dA)^2.
$$
Shifting $k\mapsto k+n$ leads to an equivalent theory, since $ndA=0$. 
 
The 5d action is gauge-invariant on a closed 5-manifold, but on a manifold with a boundary $X$ its variation under $A\mapsto A+df$ is
$$
\frac{k}{(2\pi)^2}\int_X f (dA)^2.
$$
If we do not wish to break the $\ZZ_n$ symmetry spontaneously, we cannot introduce a 4d scalar on $X$ with an axion coupling to cancel this variation. However, by analogy with 3d, we can cancel the variation by considering a topological 4d $\ZZ_n$ gauge theory with an action
$$
S'_{4d}=\frac{n}{2\pi}\int_X b da-\frac{n}{2\pi}\int_X b dA-\frac{1}{(2\pi)^2}\int_X \phi dA da.
$$
and a transformation law
$$
a\mapsto a,\quad b\mapsto b+f \frac{dA}{2\pi}.
$$
One can check that the variation of this 4d action cancels the boundary term in the variation of the 5d action. The $\ZZ_n$ symmetry is not broken on the boundary in this case. To see it more clearly, one can dualize $b$ to a $2\pi$-periodic scalar $\psi$. It can be easily seen that $\psi$ has charge $n$ both  under the boundary $U(1)$ and bulk $U(1)$. Thus it is neutral under the $\ZZ_n$ symmetry. 
 
Alternatively, since the 5d action is the Chern-Simons action, we could have canceled the anomaly by a system of free Weyl fermions with $\ZZ_n$ charges $Q_i\in\ZZ_n$ such that
$$
\sum_i Q_i^3=k\ {\rm mod}\ n. 
$$
If one turns on interactions between the fermions which preserve $\ZZ_n$ symmetry, the system might develop a mass gap. However, if $\ZZ_n$ is not broken spontaneously, the 't Hooft anomaly matching argument guarantees that the low-energy effective theory must be a nontrivial TQFT.  

It is interesting to consider the case of a more general global symmetry $G$. For simplicity, let $G$ have the form $G_1\times\ldots\times G_N$, where each factor is isomorphic to $\ZZ_n$. In that case the continuum action for the 5d DW gauge theory can be written in terms of $U(1)$ gauge fields $A_1,\ldots,A_N$ and charge-$n$ periodic scalars $\phi_1,\ldots,\phi_n$ satisfying the constraints $d\phi_i=n A_i$. The most general action is
\begin{multline}\label{S5d}
S_{5d}=\frac{1}{(2\pi)^2} \sum_{i\leq j\leq k} K^{ijk} \int_Y A_i dA_j dA_k +\frac{1}{(2\pi)^3} \sum_{i<j<k} L^{ijkl} \int_Y d\phi_i d\phi_j A_k dA_l+\\
+\frac{1}{(2\pi)^4} \sum_{i<j<k<l<m} M^{ijklm}  \int_Y A_i d\phi_j d\phi_k d\phi_l d\phi_m. 
\end{multline}
The action density is a well-defined DB 5-cocycle provided all coefficients $K^{ijk}, L^{ijkl}, M^{ijklm}$ are integral. On a closed 5-manifold $\exp(iS_{5d})$ is invariant under the gauge transformation $A_i\mapsto A_i+df_i$. Without loss of generality one may assume that $K^{ijk}$ is completely symmetric in all three indices, $M^{ijklm}$ is completely anti-symmetric, while $L^{ijkl}$ is anti-symmetric in the first three indices, and rewrite the action as follows:
\begin{multline}\label{S5d}
S_{5d}=\frac{1}{(2\pi)^2} \sum_{i\leq j\leq k} K^{ijk} \int_Y A_i dA_j dA_k +\frac{1}{(2\pi)^3} \sum \frac{1}{3!} L^{ijkl} \int_Y d\phi_i d\phi_j A_k dA_l+\\
+\frac{1}{(2\pi)^4} \sum \frac{1}{5!}M^{ijklm}  \int_Y A_i d\phi_j d\phi_k d\phi_l d\phi_m. 
\end{multline}

Note that only the first  term in (\ref{S5d}) (the one proportional to $K^{ijk}$) has the Chern-Simons form. This means that in general an 't Hooft anomaly for  a symmetry group $G=\ZZ_n^N$ cannot be computed by embedding each of the $\ZZ_n$ factors into a $U(1)$, computing the anomaly for $U(1)^N$ and reducing modulo $n$. On the other hand, if the anomaly arises from free Weyl fermions, and the symmetry group acts linearly on the fermions, $G$ is naturally a subgroup of $U(M)$ for some integer $M$, and it is well-known that the anomaly in this case is proportional to the Chern-Simons action. This implies that there exist 4d anomalies which cannot be realized by free fermions and thus require interactions. Another way to explain the distinction between free fermion anomalies and general anomalies is to say that free fermion anomalies are cubic, while the general anomaly can be also contain quartic and quintic terms. 

Let us give an example of a 4d theory with a quintic 't Hooft anomaly. Consider a topological gauge theory in 4d with gauge group $\ZZ_n^{N(N-1)/2}$. 
In the continuum it is described by $N(N-1)/2$ $U(1)$ gauge fields $a_{ij},$ where $i<j$ and $i$ and $j$ take values from $1$ to $N$, and the same number of Lagrange multiplier fields $b_{ij}$ which are B-fields (i.e. a DB 2-cocycles). It is convenient to define $a_{ij}, b_{ij}$ for all $i,j$ so that $a_{ij}=-a_{ji},$ $b_{ij}=-b_{ji}$, and $a_{ii}=0$, $b_{ii}=0$. We postulate the following transformation law for the fields under the gauged $\ZZ_n^N$ symmetry:
$$
a_{ij}\mapsto a_{ij}-f_i \frac{d\phi_j}{2\pi}+f_j\frac{d\phi_i}{2\pi}, \quad b_{ij}\mapsto b_{ij}-\frac{1}{2 (2\pi)^2} \sum_{k,l,m} N^{ijklm} f_k d\phi_l d\phi_m,
$$
where $N^{ijklm}$ is a completely anti-symmetric integral tensor. Consider now the following action:
\begin{multline}
S_{4d}=\frac{n}{4\pi}\sum_{i,j} \int_X  a_{ij}db^{ij}+\frac{1}{(2\pi)^3} \sum_{i,j,k,l,m} \frac{1}{2\cdot 3!}N^{ijklm} \int_X a_{ij} d\phi_k d\phi_l d\phi_m+\\
+\frac{1}{(2\pi)^2}\sum_{i,j} \frac12 \int_X b_{ij} d\phi_i d\phi_j.
\end{multline}
Its gauge variation is independent of $a$ and $b$ and has the form
$$
-\frac{1}{(2\pi)^4}\sum_{i,j,k,l,m} \frac{5}{12} N^{ijklm} \int_X f_i d\phi_j d\phi_k d\phi_l d\phi_m.
$$
It can be canceled by the gauge variation of the bulk action (\ref{S5d}) with $K=L=0$ and $M^{ijklm}=10 N^{ijklm}$. Again the total symmetry group is a noncommutative extension of $\ZZ_n^N$ by $\ZZ_n^{N(N-1)/2}$, and in the presence of the coupling of $a$ to $(d\phi)^3$ this leads to an 't Hooft anomaly. 

We expect that just as in 3d one can construct examples of 4d bosonic theories with 't Hooft anomalies which cannot be canceled by anomaly inflow from a DW theory in 5d. We leave a detailed investigation of such theories to future works.

\section*{Acknowledgments} A. K. would like to thank P. Etingof and V. Ostrik for discussions. R. T. would like to thank X. Chen, A. Henriques, and A. Takeda for discussions. We are especially grateful to V. Ostrik for communicating to us some unpublished results of V. Drinfeld. The work of A. K. was supported in part by the DOE grant  DE-FG02-92ER40701.

\end{document}